\documentclass[%
 reprint,
 amsmath,amssymb,
 aps,
 showkeys
]{revtex4-1}

\usepackage{graphicx}
\usepackage{dcolumn}
\usepackage{bm}
\usepackage{stmaryrd}
\usepackage{enumitem}
\usepackage[table,xcdraw]{xcolor}
\usepackage[english]{babel}

\usepackage{xr}

\DeclareMathAlphabet\mathbfcal{OMS}{cmsy}{b}{n}

\begin{document}

\preprint{APS/123-QED}

\title{Estimating the outcome of spreading processes on networks with incomplete information: a mesoscale approach}


\author{Anna Sapienza}
\affiliation{USC Information Sciences Institute, CA, USA}
\affiliation{Data Science Laboratory, ISI Foundation, Turin, Italy}

\author{Alain Barrat}
\affiliation{Aix Marseille Univ, Universit\'e de Toulon, CNRS, CPT, Marseille, France}
\affiliation{Data Science Laboratory, ISI Foundation, Turin, Italy}

\author{Ciro Cattuto}
\author{Laetitia Gauvin}

\affiliation{Data Science Laboratory, ISI Foundation, Turin, Italy}

\date{\today}

\begin{abstract}

Recent advances in data collection have facilitated the access to time-resolved human proximity data that can conveniently be represented as temporal networks of contacts between individuals. While this type of data is fundamental to investigate how information or diseases propagate in a population, it often suffers from incompleteness, which possibly leads to biased conclusions.  A major challenge is thus to estimate the outcome of spreading processes occurring on temporal networks built from partial information. To cope with this problem, we devise an approach based on Non-negative Tensor Factorization (NTF) -- a dimensionality reduction technique from multi-linear algebra. The key idea is to learn a low-dimensional representation of the temporal network built from partial information,
to adapt it to take into account temporal and structural heterogeneity properties known to be crucial for spreading processes occurring on networks, and to construct
in this way a surrogate network similar to the complete original network.

To test our method, we consider several human-proximity networks, on which we simulate a loss of data. Using our approach on the resulting partial networks, we build a surrogate version of the complete network for each. We then compare the outcome of a spreading process on the complete networks (non altered by a loss of data) and on the surrogate networks. We observe that the epidemic sizes obtained using the surrogate networks are in good agreement with those measured on the complete networks. Finally, we propose an 
extension of our framework when additional data sources are available to cope with the missing data problem. 
\end{abstract}



\maketitle

\section{Introduction}

The advances made in data collection technologies have led to a wealth of high-resolution time-resolved data. Mobile sensing devices, social networking applications and wearable sensors have indeed significantly contributed to monitor social interactions and physical proximity of individuals in time~\cite{lane2010survey,miluzzo2008sensing,eagle2006reality,cattuto2010dynamics,stopczynski2014measuring,salathe2010high}. 
Such fine-grained data monitoring is crucial for a deeper study of human proximity dynamics described by complex temporal networks -- in which links are drawn between nodes representing individuals when they are in close-range~\cite{T1} -- and their interplay with contagion processes. Physical proximity interactions play indeed 
a fundamental role in conveying information or in the spread of diseases~\cite{read2008dynamic,obadia2015detailed,voirin2015combining,he2010message}. 
They can thus inform our understanding of how messages or infectious diseases such as flu-like illnesses propagate among individuals. 

However, despite the efforts made to increase the accuracy in the data collection, relational data often suffer from incompleteness resulting in missing links in empirical networks~\cite{schafer2002missing}. This lack of information can arise for several reasons: limited participation during surveys, incomplete records (diary-based, device-based)~\cite{smieszek2011collecting,ouzienko2014imputation,mastrandrea2016estimate,smieszek2016contact}, and technical issues occurring during the data collection process. In the case of smartphone sensing such as in \cite{eagle2006reality,stopczynski2014measuring}, proximity might be undetected during some time windows. For instance, people might turn off their Bluetooth or Communication Data Records might not provide access to the vicinity of individuals to a cell tower at each time, possibly leading to undetected co-presence events. In the case of self-reporting of sexual relationships, individuals might choose not to disclose all of their partners, which leads to biases when focusing on the spread of sexually transmitted disease \cite{ghani1998sampling}. Data incompleteness can affect the measured properties and structure of temporal networks~\cite{kossinets2006effects,rocha2017sampling} and can reflect on the simulated evolutions of contagion processes, leading to inaccurate conclusions~\cite{genois2015compensating,vestergaard2016impact}. The investigation of information or disease propagation processes 
on networks built from such data must thus be undertaken carefully.

Several approaches have been put forward to cope with missing links in networks~\cite{huisman2009imputation}. 
These methods include: distributional models, which estimate the likelihood of the presence of a link on the basis of the observed links and nodes attributes~\cite{guimera2009missing}, hierarchical structure methods~\cite{clauset2008hierarchical}, stochastic block models~\cite{airoldi2008mixed}, and expectation maximization methods~\cite{kim2011network}, which try to extract the connectivity patterns in the available part of the network to infer and complete the unknown part.
However, the goal of these methods is the exact recovery of single links, a complicated task that becomes 
nearly impossible to achieve when a large amount of data is missing. Moreover, this might actually not be necessary if the goal is to estimate the
global outcome of a contagion process at the population level and not the risk concerning a specific individual. 
Other approaches have been developed  to specifically estimate important properties of epidemic spread  and information cascade without trying to recover the original network~\cite{sadikov2011correcting,genois2015compensating,mastrandrea2016estimate,fournet2017estimating,ghonge2017inferring}.
These methods are however either process specific or rely on the existence of known mesocale structures (such as groups in the population) in the network, together with the knowledge of the structure to which each population member belongs.

Here, we propose a self-contained approach that does not rely on such a priori knowledge. As in~\cite{genois2015compensating,mastrandrea2016estimate}, the aim of this work is not to recover the exact links which are missing in the data but instead to build a surrogate version of the network of interactions. 
The main difference with these previous works, that is a building block of our approach, 
is that we uncover {\it in the incomplete data} mesoscale structures that involve nodes affected by missing activity. To study the network at the mesoscale level, we take advantage of tensor decomposition techniques, already applied in several fields~\cite{P1}, to extract both the topological and temporal properties of the network~\cite{P2}. 
An important aspect of our method is that it does not rely on the availability of metadata or external information on the nodes, and that it 
allows us to recover fundamental properties of the studied temporal network, such as the temporal activity of nodes with partial information (i.e., the
temporal evolution of their number of contacts). We use this approach to build a surrogate version of the network that yields a correct estimate of
the outcome of a simulated spreading process occurring on the network. 
Furthermore, we extend our method to take advantage of other sources of information that might be available, such as information deriving from subsidiary data sources.

The paper is organized as follows: in Section~\ref{Not} we present the notations used in the paper; in section~\ref{PS} we describe the problem statement; in Section~\ref{method} we explain the method that we developed to carry out the study; in Section~\ref{Res} we report the results achieved by our approach in the study of several temporal human proximity networks; in Section~\ref{Con} we discuss the performance and limitations of the method and future research directions.

\section{Notations}
\label{Not}
The following notations are used throughout the present paper. Lower-case letters denote scalar variables, e.g., $t$, capital letters denote defined constants, e.g., $T$, and boldface lower-case letters denote vectors, e.g., $\mathbf{t}$.

Matrices are denoted by boldface capital letters, e.g., $\mathbf{T}$, where the $i$-th column of a matrix $\mathbf{T}$ is $\mathbf{t}_i$ and the $\left(i,j\right)$-entry is $t_{ij}$. Third order tensors are denoted by bold calligraphic letters, e.g., $\mathbfcal{T}$, whose $\left(i,j,k\right)$-entry is $t_{ijk}$. 

The tensor product is denoted by $\circ$, the Hadamard (element-wise) product by $\ast$, the Kronecker product by $\otimes$, and the outer product by $\cdot$. $\|\cdot \|_F$ stands for the Frobenius norm.
\section{Problem Statement}
\label{PS}

The purpose of the present work is the development of a methodology able to reproduce the outcome of contagion processes on temporal networks of human
interactions, starting from incomplete information on these networks. 
We consider a scenario in which part of the activity of a fraction of the network nodes (i.e. part of their interactions over time) is missing in the data. 
To provide an estimate of the outcome of spreading processes on the network, we do not try to recover the exact missing links and interaction events of these nodes. 
Our method aims instead at building a surrogate version of the complete network by taking advantage of the only information that is available. 
As we will describe in details in Sec.~\ref{approxntf} and Sec.~\ref{approxjntf}, this information can be related either
only to the partial network of human contacts or to richer data than can be used as a proxy for human proximity. 
For instance, we could have access both to the partial temporal contact network and to the approximated location of individuals provided by smartphones, through the GPS, 
Bluetooth or WiFi signals: such additional information can conveniently 
be represented as a temporal bipartite network between individuals and locations, in which a link is drawn between an individual
and a location when the individual is detected in that location. 
We will show how such information can be integrated in our framework.

The underlying assumption of our method is that we can leverage the mesoscale properties of the partial network, 
such as the presence of correlations in the node activities, to build a surrogate version of the complete network. To extract these mesoscale properties from the incomplete data we rely on the Non-negative Tensor Factorization (NTF) technique~\cite{P1,*bro1997fast}. In particular, our method is based on a NTF framework handling missing values~\cite{acar2010scalable,royer2012nonnegative}. 
By applying the NTF on a tensor representing a temporal network we can indeed identify groups of nodes having similar connectivity patterns and whose links have similar activation times. Each of these groups can be seen as a sub-network. By studying the structure and the temporal activity of each sub-network, we can infer the properties of the nodes whose activity is partially missing.   

However, standard tensor decomposition techniques such as NTF are based on the assumption that noise in the data follows a Gaussian distribution. Thus, even if the data are intrinsically heterogeneous, as is typically the case in human contact data \cite{barrat2015face}, the decomposition will provide a low-rank approximation in which these heterogeneous characteristics are not preserved~\cite{hayashi2010exponential}. 
As a result, the NTF tends to approximate the network in a way that makes the properties of the sub-networks extracted more homogeneous than in the original network. This trend towards homogeneity changes the temporal and structural properties of the network such as the distribution 
of the number of contacts per link (i.e., the number of activation times of each link) in the aggregated network, and thus has to be carefully taken into account. Indeed, heterogeneity, both in the aggregated number or duration of contacts per link and in temporal properties (e.g., broad inter-event time distributions) have 
been shown to play a key role in spreading processes occurring over static and temporal networks~\cite{vazquez2007impact,iribarren2009impact,smieszek2009mechanistic,Stehle2011,machens2013infectious,gauvin2013activity,vestergaard2014memory}. 

Since we are interested in recovering the outcome of such processes, we need to preserve the heterogeneity properties of the network. Thus, instead of directly using the approximated network provided by the NTF, we will use the information about the nodes belonging to each sub-network and their related activity patterns as a guide to build a surrogate network. In this surrogate network we will reintroduce the heterogeneity properties as described below.

Our method to build a surrogate network from partial information and recover the outcome of contagion processes is thus divided in two main steps:
\begin{enumerate}[label=\Alph*]
\item.\label{A} The extraction of mesoscale structures from the partial temporal network through NTF adapted to handle incomplete information; 
\item. the construction of a surrogate network, in which we use both the information provided by step \ref{A} and the heterogeneity properties of the network.
\end{enumerate}

\section{Method}
\label{method}
\subsection{Extracting mesoscale structures from a partial temporal network}
\label{approxntf}
Three-way tensors (i.e. three-dimensional arrays) are natural representations of temporal networks: given an undirected temporal network, composed by $N$ nodes and $k =1,\dots,K$ time intervals, we can represent its snapshots  $G_k = (V,E_k)$, which have $|V|=N$ nodes and set of links $E_k$, by $K$ adjacency matrices $\mathbf{M}_k\in\mathbb{R}^{N\times N}$ of the form:
\[
\mathbf{M}_k = 
\begin{cases}
m_{ij} = 1 \;\; \mbox{if} \;\;\left(i,j\right) \in E_k\,, \\
m_{ij} = 0 \;\; \mbox{otherwise}\,.
\end{cases}
\]
These adjacency matrices form the slices of a tensor $\mathbfcal{T}\in\mathbb{R}^{I\times J\times K}$, with here $I=J=N$. In the case of missing data, 
zero values in the tensor can either correspond to no activity or to undetected activity. We however need to factorize 
only the part of the tensor that corresponds either to measured activity or to actual inactivity (absence of contact). 
To this aim, the zero entries that correspond to possibly undetected activities have to be masked in the tensor. 
If a node with partial information has no activity at all measured during a given snapshot (in a given slice of the tensor), 
we consider that the related zero entries in the tensor $\mathbfcal{T}$ might correspond to possibly undetected activities, i.e., possibly missing contacts. 
On the other hand, if a node, for which we know that only partial information is present, 
has at least one contact measured in a slice, we assume that no information was lost at all in that tensor slice because of that node.
In other terms, we assume
for each node and each time slice that either all or none of the activity recorded by that node in that time slice is present in the data.
This would be for instance the case if a sensor measuring proximity is turned off or if the GPS coordinates of a mobile user are not collected during a time window:
all the proximity information concerning this sensor in this time window is then lost. However,
if the measuring device is not turned off, all the proximity relationships with other devices that are also on during that time window are present in the incomplete data.

We thus introduce a binary tensor $\mathbfcal{W}$, of the same size of $\mathbfcal{T}$, whose entries are defined as
\[
w_{ijk} = 
\begin{cases}
0 \;\; \mbox{if} \;\; \mbox{$i$ or $j$ has no activity in the time window $k$,}\; \\
1 \;\; \mbox{otherwise.} \,
\end{cases}
\]
This tensor is used to mask the part of the tensor $\mathbfcal{T}$  that might be linked to possibly undetected activity.
The approximation of the masked tensor $ \mathbfcal{T}\ast\mathbfcal{W}$ consists in minimizing the following cost function with non-negative constraints~\cite{royer2012nonnegative}:
\begin{align*}
&f_w\left(\boldsymbol\lambda,\mathbf{A}, \mathbf{B}, \mathbf{C}\right) 
= \left\|\mathbfcal{W}\ast\left(\mathbfcal{T} - \sum^{R}_{r=1}\lambda_r \mathbf{a}_r\cdot  \mathbf{b}_r \cdot \mathbf{c}_r\right)\right\|^2_F \;,
\end{align*}
where $R$ is the rank of the approximation and is hereinafter called the number of components. 
The vectors $\mathbf{a}_r$, $\mathbf{b}_r$, and $\mathbf{c}_r$ with $r\in \left[1,R\right]$ form
respectively the factor matrices $\mathbf{A}\in\mathbb{R}^{I\times R}$, $\mathbf{B}\in\mathbb{R}^{J\times R}$, $\mathbf{C}\in\mathbb{R}^{K\times 
R}$. Each tuple of vectors $(\mathbf{a}_r,\mathbf{b}_r, \mathbf{c}_r)$ describes a mesocale strucure. As we explain below, $\mathbf{a}_r$ and $\mathbf{b}_r$ indicate which
nodes and links participate to the mesoscale structure $r$, while $\mathbf{c}_r$ 
describes the temporal activity of the structure $r$. 
For the sake of readability, the cost function is re-written in the following form:
\begin{align}
\label{eqm}
&f_w\left(\boldsymbol\lambda,\mathbf{A}, \mathbf{B}, \mathbf{C}\right) 
= \|\mathbfcal{W}\ast\left(\mathbfcal{T} - \llbracket\boldsymbol\lambda; \mathbf{A}, \mathbf{B}, \mathbf{C}\rrbracket\right)\|^2_F .
\end{align}

Using the mask $\mathbfcal{W}$ amounts to approximating the tensor $\mathbfcal{T}$ based only on information of which we are certain: 
we are approximating only the part of the tensor composed of elements that are either $1$ corresponding to measured events or $0$ corresponding
to real absence of contact, without taking into account all the $0$ values that might correspond to undetected activity.

In other terms, minimizing the cost function $f_w$ corresponds to finding the best approximation 
of the non-masked part of the tensor by a sum of components  corresponding each to a mesoscale structure. 
The analysis of $ \mathbf{A}, \mathbf{B}$ and $\mathbf{C}$ yields then information on which link is involved in which 
mesocale structure, including in particular the links involving nodes with incomplete information.
Several methods are available to estimate the factor matrices $\mathbf{A}, \mathbf{B}$ and $\mathbf{C}$~\cite{kim2007,blockprp}, and
details are given in Section~\ref{methods}.

\subsection{Extracting mesoscale structures from coupled temporal networks}
\label{approxjntf}

Here, we propose an extension of the mesoscale structure detection method to the case in which we have access to richer information -- not expressible in one single temporal network. The integration of such information might help to better recover the missing entries in the tensor describing the temporal contact network: 
for instance, if we have access, in addition to the partial contact network, 
to the location of individuals involved in the contacts, the latter can contribute to recover the missing information on contacts.
To integrate such additional data, we propose to use the so-called Joint Non-negative Tensor Factorization (JNTF)~\cite{acar2011all} that
makes it possible to decompose multiple temporal networks at once in a coupled manner (in practice,  
 we will consider the partial temporal contact network and a position network evolving with time). 

Let us consider the general case of $S$ different temporal networks, each represented by a tensor $\mathbfcal{T}_s$, with possibly different dimensions as they might represent different types of information. We can approximate them in a coupled way by computing the following minimization problem with non-negative constraints:
\begin{align}
\label{jntfS}
\mbox{min}&\sum^S_{s=1}\|\mathbfcal{T}_s - \llbracket \boldsymbol\lambda_s; \mathbf{A}_s, \mathbf{B}_s, \mathbf{C}_s \rrbracket\|^2_F \\
&\mbox{s.t.}\;\; \boldsymbol\lambda_s, \mathbf{A}_s, \mathbf{B}_s, \mathbf{C}_s\geq 0 \nonumber\,.
\end{align}
Different couplings can be considered by imposing that some of the factor matrices $\mathbf{A}_s, \mathbf{B}_s, \mathbf{C}_s$ 
in the equation are equal for different values of $s$. 
The idea behind the introduction of this coupling is that different networks can provide partially redundant information, and that this redundancy is 
relevant for recovering missing entries. For instance, if we have access to the locations of nodes but only to a partial information concerning their contacts, 
we can couple the decomposition of the tensor representing the partial contact network to the decomposition of the tensor
representing the time evolution of the location of nodes, and gain in this way information on the possible contacts over time. 
In practice, we would impose $\mathbf{A}_1=\mathbf{A}_2$ and $\mathbf{C}_1=\mathbf{C}_2$, which correspond respectively
to imposing the same nodes' memberships and the same activity timeline for each mesoscale structure in the resulting approximations of the 
contact and location tensors.

The joint factorization of tensors -- including one with missing information $\mathbfcal{T}_{s'}$ --, can be adapted to handle missing values in the same way as the NTF:
\begin{eqnarray*}
\min&\Big( \alpha_{s'}\|\mathbfcal{W}\ast\left(\mathbfcal{T}_{s'} - \llbracket\boldsymbol\lambda_{s'}; \mathbf{A}_{s'}, \mathbf{B}_{s'}, \mathbf{C}_{s'}\rrbracket\right)\|^2_F \\
+ & \sum^S_{\substack{s=1 \\ s\neq s'}}\alpha_{s}\|\mathbfcal{T}_s-\llbracket \boldsymbol\lambda_s; \mathbf{A}_s, \mathbf{B}_s, \mathbf{C}_s\rrbracket\|^2_F \Big)\\
&\mbox{s.t.}\;\; \boldsymbol\lambda_s, \mathbf{A}_s, \mathbf{B}_s, \mathbf{C}_s\geq 0 \;\forall s\nonumber\,.
\end{eqnarray*}
To solve this minimization problem we adapted the active-set-like method with Karush-Kuhn-Tucker optimality conditions ~\cite{park2012,kim2007} 
(see Section~\ref{methods} for details). 
In the following, we set  all the $\alpha_s$ parameters equal to $1$. We note however that they could 
be used to tune the relevance of the information provided by each tensor. 
As an example, by setting $\alpha_{s'}>\alpha_{s}$, $\forall s \ne s'$ we would give more importance to the 
information provided by the network with partial information than to the one given by the subsidiary data.

\subsection{Surrogate network}
\label{netrec}

Both types of factorization, either from a partial network or from coupled temporal networks, provide approximations $\mathbfcal{T}_{app}$ 
of the temporal networks as sums 
of components, each corresponding to a mesoscale structure. Each mesoscale structure is fully 
described by the respective columns of  $\mathbf{A}$, $\mathbf{B}$ and $\mathbf{C}$ and
can be seen as a sub-network composed of links with similar temporal properties, as we now describe.

To analyze the mesoscale structures revealed by the factorization step and interpret them as sub-networks, we estimate the memberships of the links 
using $\mathbf{A}$ and $\mathbf{B}$ and we determine the activation times by binarizing $\mathbf{C}$. 
First, the membership weight of each link $\left(i,j\right)$ to the $r$-th component is given by the  $\left(i,j\right)^{th}$ element of the following Kronecker product:
\[
\mathbf{a}_r\otimes\mathbf{b}_r = \mathbf{a}_r\cdot\mathbf{b}^T_r .
\]
As we consider undirected networks, the actual membership of each link is symmetrized in the following way:
\[
\frac{\mathbf{a}_r\cdot\mathbf{b}^T_r+  \mathbf{b}_r\cdot\mathbf{a}^T_r}{2} .
\]
We consider that the links with the largest membership weights, composing $95\%$ of the total sum of the squared memberships,  
belong to the component $r$ (note that, depending on the membership values, a link could belong to more than one component). 
Then, to detect the times in which each mesoscale structure is active, we consider the matrix $\mathbf{C}$, which summarizes the temporal activity of each component:
by using the Otsu method~\cite{SD1}, a common way to perform binary thresholding, 
we transform the temporal activity of each component in its binary version ($1$ if it is active, $0$ otherwise). 

These two steps allow us to determine for each mesoscale structure when it is active and which links it involves.
For each structure, we then select the links involving at least one node with partial information and the times in which 
its activity was potentially lost (in the sense that no activity of that node was recorded at all for this particular time), and we add the corresponding
elements of the structure to the partial network $\mathbfcal{T}$ to create the surrogate tensor  $\mathbfcal{T}_{surrogate}$. 
The rationale is that the factorization has allowed us to determine, for the links for which only partial information is available,
to which mesoscale structures they belong. Using the activity timelines of the mesoscale structures, we thus reconstruct the missing parts of the activity timelines
of these links. 
Moreover, the binarization step enables to reintroduce the heterogeneity in the network. As mentioned previously, the properties of the sub-networks extracted tend to be more homogenous, in particular the distribution
of weights of each sub-network is more homogenous than in the real data. 
Indeed, the factorization finds groups of links that have strongly correlated activity timelines and approximates them as fully correlated. 
As a result, the distribution of the link weights in the network described by $\mathbfcal{T}_{app}$ -- the number of times in which each link is active -- 
becomes more homogeneous and the activity tends to be less bursty than in the empirical network.
By binarizing the approximation in the way described above we are removing links which were associated to 
small weights and increasing weights of links with the highest weights in each component of the approximated case. 
This contributes to make the overall distribution of weights more heterogenous.

When the JNTF is used, an additional step turns out to be relevant.
The JNTF indeed yields components influenced by both the activity in the partial proximity network and the subsidiary temporal network 
(such as a network linking individuals and locations). Thus, the JNTF will approximate the network in a way
based also on co-presence events of individuals. As co-presence is a necessary but not sufficient condition for two individuals to be in contact, the links added
when creating  $\mathbfcal{T}_{app}$ are less likely to correspond exactly to contacts that were missing than in the NTF case. 
Moreover, for the same reason the number of links in $\mathbfcal{T}_{app}$ and their weights might be greater than in the original data and they might present
less bursty activity patterns.
As mentioned in~\ref{approxjntf}, the weight of the information provided by the second tensor can of course be tuned in the JNTF
by adding coefficients so that the components extracted by the factorization are more representative of the network with partial 
information than of the other one. The investigation of this possibility is however outside the scope of this paper.
The procedure of binarization of $\mathbfcal{T}_{app}$ described above needs thus to be completed to make 
the properties of the network more heterogeneous. 
In order to do this, we  have to discard some of the times in which the links involving nodes with partial information are active in $\mathbfcal{T}_{app}$ to compensate for the overestimation and get closer to the empirical network. 
The key idea here is to zero out some elements in $\mathbfcal{T}_{surrogate}$ to recover a weight distribution comparable to the empirical one,  
measured on the links involving only nodes with no missing information. We pick from this distribution a number of weights 
equal to the number of links having partial information, for which we have to adjust the weights, and we assign them at random to these links.
Finally, we compare for each link the new value to its weight in $\mathbfcal{T}_{surrogate}$: 
 if the new weight is smaller than the old one, we erase at random parts of the link activity that are present in $\mathbfcal{T}_{surrogate}$, 
until we reach the new weight; if instead the new weight is larger than the old one, we do not act on that link's activity. 
The reason why we can rely on the weight distribution measured on the partial network 
is due to its robustness to sampling, as shown in~\cite{cattuto2010dynamics,mastrandrea2016estimate,genois2015compensating,sadikov2011correcting} for
various sampling procedures and in the Supplementary Information S.\ref{Swd} for the sampling considered here.
After the reassignment of weights, the resulting tensor $\mathbfcal{T}_{surrogate}$  is used to approximate the whole temporal network of contacts
and perform simulations of spreading processes.

\section{Results}
\label{Res}
We apply the method described in the previous sections to three temporal human proximity networks. The data were collected  
by the SocioPatterns collaboration (www.sociopatterns.org) 
in two conferences in Italy (HT09) and France (SFHH), and in a primary school in France (LSCH). In each case,
the proximity of individuals in the network was measured with wearable sensors able to capture face-to-face contacts occurring in a $1-2$-meter range 
with a $20$-second time resolution. As for the purpose
of the present paper we do not need such a high-resolution, we aggregate the data to a $15$-minute resolution. 
The HT09 dataset was collected during the ACM Hypertext 2009 conference~\cite{szomszor2010semantics}. 
The resulting temporal network has  $N=113$ nodes and $K=237$ snapshots.
The SFHH dataset was collected during the conference of the Soci\'et\'e Fran\c caise d'Hygi\`ene Hospitali\`ere, yielding a temporal network with $N=417$ nodes and $K=129$ time snapshots~\cite{cattuto2010dynamics}. Finally, the LSCH dataset was collected in a primary school in Lyon~\cite{stehle2011high}. 
The resulting network has $N=241$ nodes and $K=130$ snapshots. 
We consider all the snapshots as unweighted networks (meaning that each element of the tensor is either $0$ or $1$).

We simulate on each dataset a loss of data determined by a fraction of nodes $p_{nodes} \in\left[0.1, 0.2, 0.4\right]$  
for which we zero out the activity occurring during the first half of the total timespan, i.e., the information concerning each of these nodes is lost over a fraction 
$p_{times}=0.5$ of the temporal snapshots. 
In practice, for each dataset we start from the complete tensor $\mathbfcal{\hat{T}}$ and 
we create a tensor with partial information $\mathbfcal{T}$ by erasing, in the first half of the temporal slices of $\mathbfcal{\hat{T}}$, the elements 
related to $N p_{nodes}$ chosen at random. 
To test the limits of our method, we also considered cases where either all the nodes lost some 
information or where a fraction of the nodes lost all their activity. More details on the cases dealt with are provided below.

\subsection{Approximated network}
For each dataset and for each data loss scenario, we performed the approximation of $\mathbfcal{T}$, i.e., 
the minimization of Eq.~\eqref{eqm}, with a number of components selected using the so called Core Consistency Diagnostic~\cite{bro2003new} as a guide (see 
Sec.~\ref{methods} for details). We perform $20$ decompositions in each case, varying the initial conditions, and we use the one 
with the highest core consistency value. Moreover, for each value of $p_{nodes}$ we repeat the procedure on $10$ different sets of 
nodes with missing information chosen at random, in order to evaluate the variations in the performance of our method.

From the factorization of the tensors with missing information
$\mathbfcal{T}_{HT09}$, $\mathbfcal{T}_{SFHH}$, and $\mathbfcal{T}_{LSCH}$, we recover a first approximated version of the complete network 
$\mathbfcal{T}_{app}$ for each dataset. 
We evaluate the results provided by the decomposition by computing the Pearson's coefficient between the complete and approximated temporal activities of the nodes 
for which only partial information is available in $\mathbfcal{T}$. The correlation is measured only on the part of the activity that is missing in the partial data.
We report the correlation coefficients found in Table~\ref{tabcoeff} and we show in the Supplementary Information
S.\ref{Sappnet} a representative example of the temporal activities of 
these nodes with partial information in the complete, partial, and approximated networks.
The correlation coefficients shown are measured on $10$ different sets of nodes with missing information for each dataset. The ranges of Pearson's coefficient indicate positive moderate to high correlation between the complete and approximated node activity, indicating a good recovery of the node temporal activities. Let us notice that by construction the approximation through the factorization relies on the existence of correlated activity patterns and consequently performs better in the presence of strong such patterns. This explains why the temporal activities are better recovered (larger Pearson correlation coefficients) in the LSCH case than in HT09 and SFHH. 
Indeed, in schools the schedule is quite constrained and all the students
of a given class have highly correlated activity timelines, leading to stronger correlated activity patterns than in conferences, during which 
attendees are free to move and interact with different people at different times.

\begin{table}[t]
\centering
\caption{Range of values of the Pearson's coefficient computed by comparing the original and approximated node activity for each node in 
the different sets considered. The table also reports the median value obtained. The p-values are lower than $10^{-3}$.}
\label{tabcoeff}
\begin{tabular}{l|c|c|c|c|}
\cline{2-5}
                           & \multicolumn{1}{l|}{$p_{nodes}$} & \multicolumn{1}{l|}{$p_{times}$} & \multicolumn{1}{l|}{Pearson's coeff.} & \multicolumn{1}{l|}{median value} \\ \hline
\multicolumn{1}{|l|}{}     & 0.1                               & 0.5                               & {[}0.53, 0.96{]}                      & 0.77                              \\
\multicolumn{1}{|l|}{LSCH} & 0.2                               & 0.5                               & {[}0.54, 0.96{]}                      & 0.76                              \\
\multicolumn{1}{|l|}{}     & 0.4                               & 0.5                               & {[}0.53, 0.96{]}                      & 0.77                              \\ \hline
\multicolumn{1}{|l|}{}     & 0.1                               & 0.5                               & {[}0.38, 0.83{]}                      & 0.54                              \\
\multicolumn{1}{|l|}{HT09} & 0.2                               & 0.5                               & {[}0.37, 0.80{]}                      & 0.52                              \\
\multicolumn{1}{|l|}{}     & 0.4                               & 0.5                               & {[}0.37, 0.73{]}                      & 0.50                              \\ \hline
\multicolumn{1}{|l|}{}     & 0.1                               & 0.5                               & {[}0.40, 0.89{]}                      & 0.57                              \\
\multicolumn{1}{|l|}{SFHH} & 0.2                               & 0.5                               & {[}0.48, 0.93{]}                      & 0.57                              \\
\multicolumn{1}{|l|}{}     & 0.4                               & 0.5                               & {[}0.40, 0.89{]}                      & 0.57                              \\ \hline
\end{tabular}
\end{table}

\subsection{Surrogate network}

As illustrated in the previous paragraphs, the approximation step achieves good results to recover single node temporal activities
by using NTF handling missing information. However, 
as discussed above, heterogeneity properties of the link weights are crucial in determining the outcome of spreading processes.
We show indeed in the Supplementary Information S.\ref{SSIRappnet} that simulating a spreading process directly on an approximated network $\mathbfcal{T}_{app}$ without
the binarization step results in a strong underestimation of the original outcome of the spreading process.
We thus apply the procedure described above and 
build a surrogate temporal network $\mathbfcal{T}_{surrogate}$  relevant to estimate the outcome of spreading processes. 
For the JNTF, for each link involving nodes with partial information, we assign a weight extracted from the distribution
of weights measured on the links with no missing information, and, if needed, we remove a part of the link's activity present in $\mathbfcal{T}_{surrogate}$
but not in $\mathbfcal{T}$ until we match this weight.

\subsection{Estimate of the outcome of spreading processes}
\label{outSIR}
To evaluate the performance of the method in estimating the outcome of spreading processes on the network, we simulate susceptible-infected-recovered (SIR) processes (see details in Sec.~\ref{sir}) over three temporal networks for each dataset: the complete, the partial, and the surrogate network. In each case we run multiple simulations 
for each set of values ($\beta$, $\mu$) of the infection and recovery probabilities per unit time. 
We focus on the couples of probabilities $\left(\beta,\mu\right)$ that satisfy the following criterion. The spreading has to be finished within the timespan of the dataset, and the epidemic size, defined as the final fraction of recovered individuals, has to be greater than $20\%$ and lower than $80\%$  (the selection is based on the median of the epidemic size). These conditions ensure to avoid the selection of parameters such that the simulations either never reach a significant epidemic size and/or are too slow with respect to the total timespan of the network. The limit for the final number of recovered individuals to $80\%$ of the entire population prevents the selection
of parameters leading to too fast spreading that are far from realistic conditions.

For each selected pair of parameter values $\beta$ and $\mu$, we compute the distribution of the epidemic sizes in the three cases (here called ``Complete", ``Partial", ``Surrogate"). As we simulate a loss of data by considering $10$ different sets of randomly chosen nodes with partial information, 
we report the median distribution of the epidemic size on the simulations over these $10$ cases as well as the $25$-th and $75$-th percentiles.
We first consider the case in which the only available information is the partial temporal network $\mathbfcal{T}$, and then 
a case in which an additional source of information is available, namely the location of individuals over time.

\subsubsection{Using only the partial temporal network}

\begin{figure*}
  \includegraphics[width=.9\textwidth]{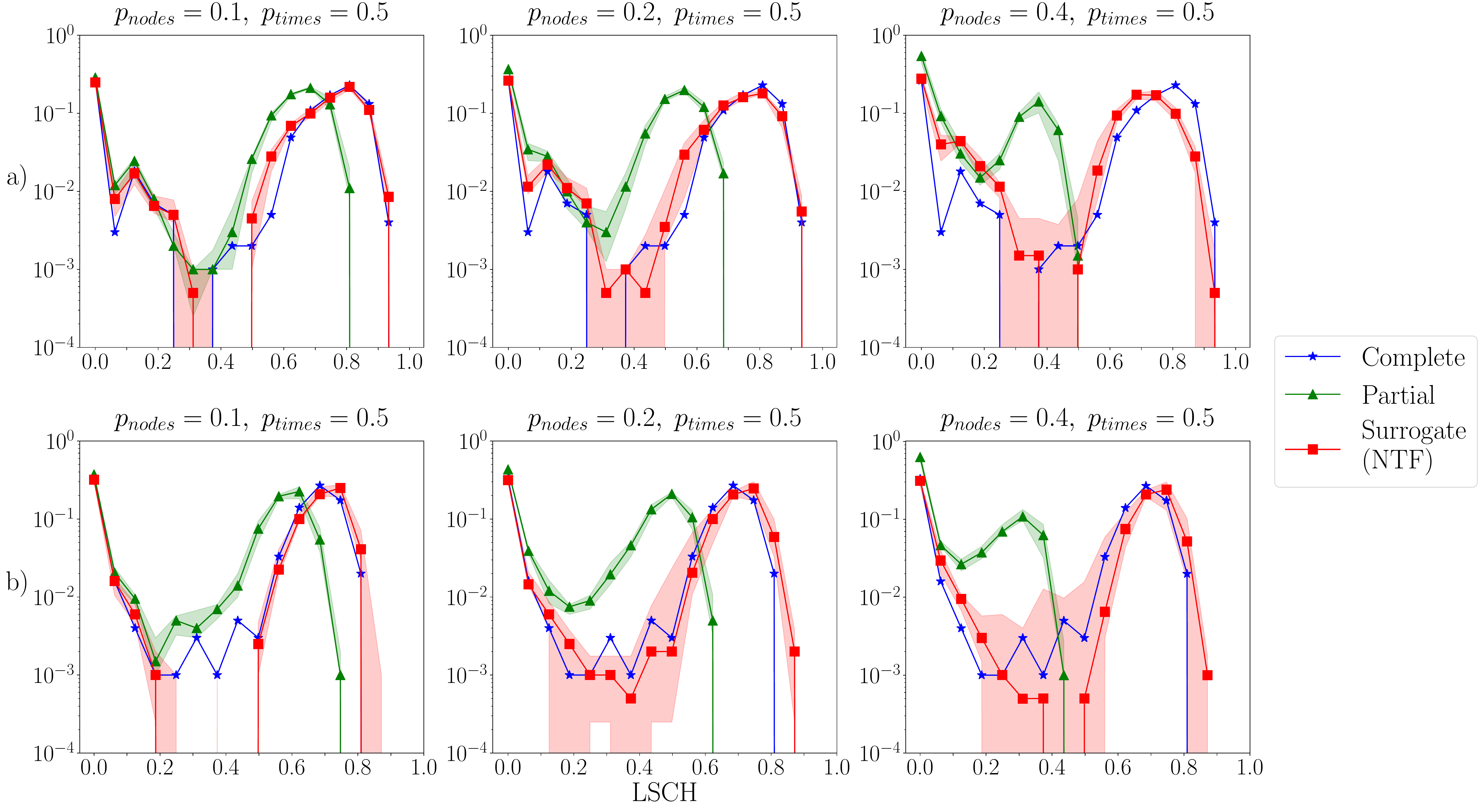}
  \caption{Distributions of epidemic sizes computed in the complete, partial, and surrogate cases for the LSCH dataset. Each panel corresponds to one
  value of the fraction $p_{nodes}$ of nodes with partial information and one couple of spreading parameters: a) $\beta = 0.3$ and $\mu = 0.3$; b) $\beta = 0.15$ and $\mu = 0.25$. 
  For the partial and surrogate cases, the symbols and 
  lines show the median distribution of the epidemic size computed from the results relative to the $10$ different sets of nodes with partial information, 
  while the shaded area is delimited by the $25$-th and $75$-th percentiles.}
\label{sirsch}
\end{figure*}

In Fig.~\ref{sirsch} we report the epidemic size distributions obtained for the LSCH dataset, for two couples of selected spreading
parameters: a) $\beta = 0.3$, $\mu = 0.3$ and b) $\beta = 0.15$, $\mu = 0.25$ and for different fractions of nodes $p_{nodes}$ with partial information. The figure shows that our method yields a good approximation
of the distribution obtained in the original case, while the one obtained by simply simulating the SIR process on the network with partial information strongly underestimates 
the epidemic size. This underestimation becomes stronger as $p_{nodes}$ increases while the estimation given by the surrogate network remains
of good quality. 

\begin{figure*}
  \includegraphics[width=.9\textwidth]{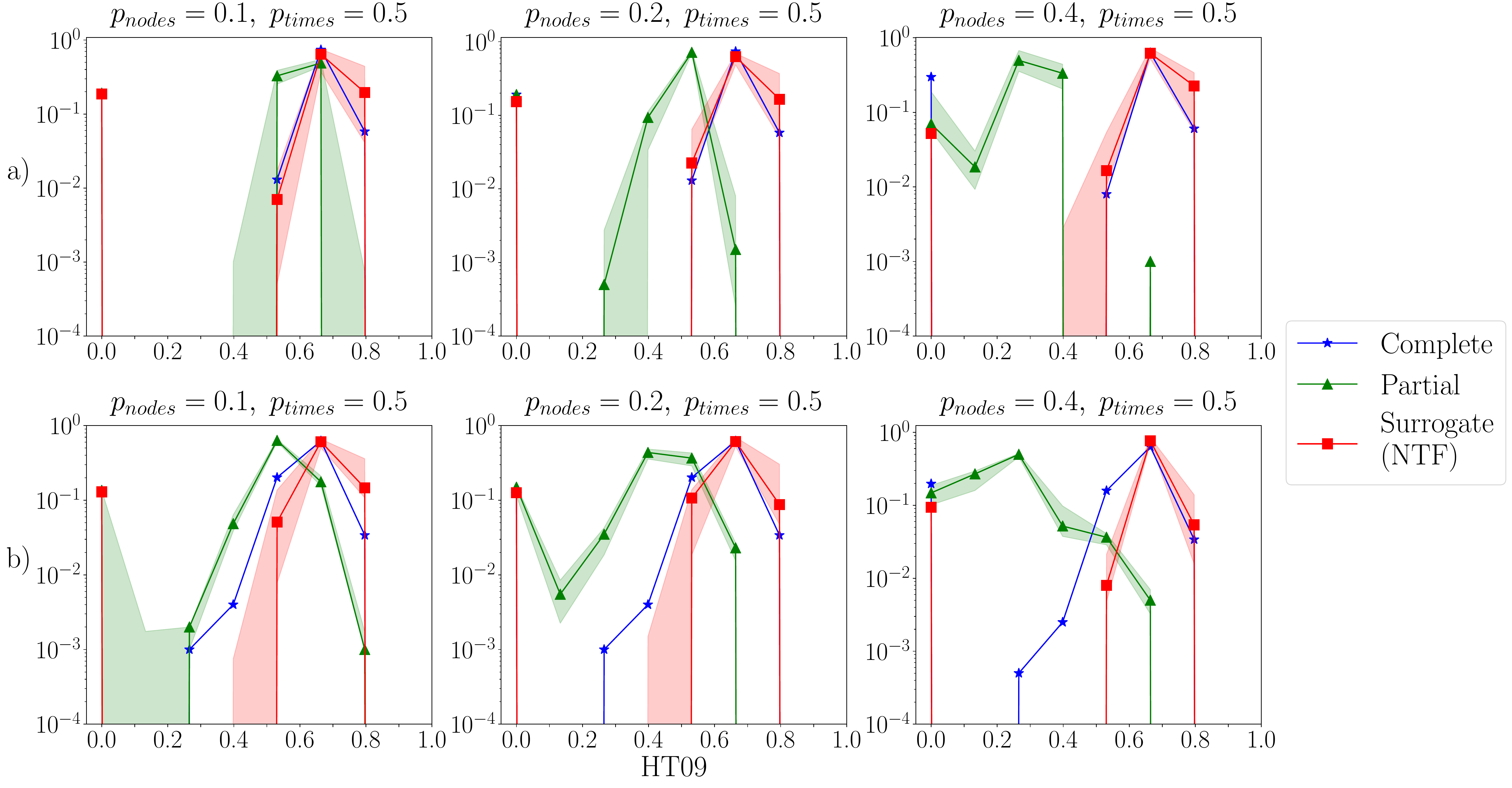}
  \caption{Distributions of epidemic sizes computed in the complete, partial, and surrogate cases for the HT09 dataset. Each panel corresponds to one
  value of the fraction $p_{nodes}$ of nodes with partial information and one couple of 
  spreading parameters: a) $\beta = 0.60$ and $\mu = 0.10$; b) $\beta = 0.25$ and $\mu = 0.05$. 
  The symbols and lines show the median distribution of the epidemic size computed from the results relative to the $10$ different sets of nodes with partial information, while the shaded area is delimited by the $25$-th and $75$-th percentiles.}
\label{sirht09}
\end{figure*}

\begin{figure*}
  \includegraphics[width=\textwidth]{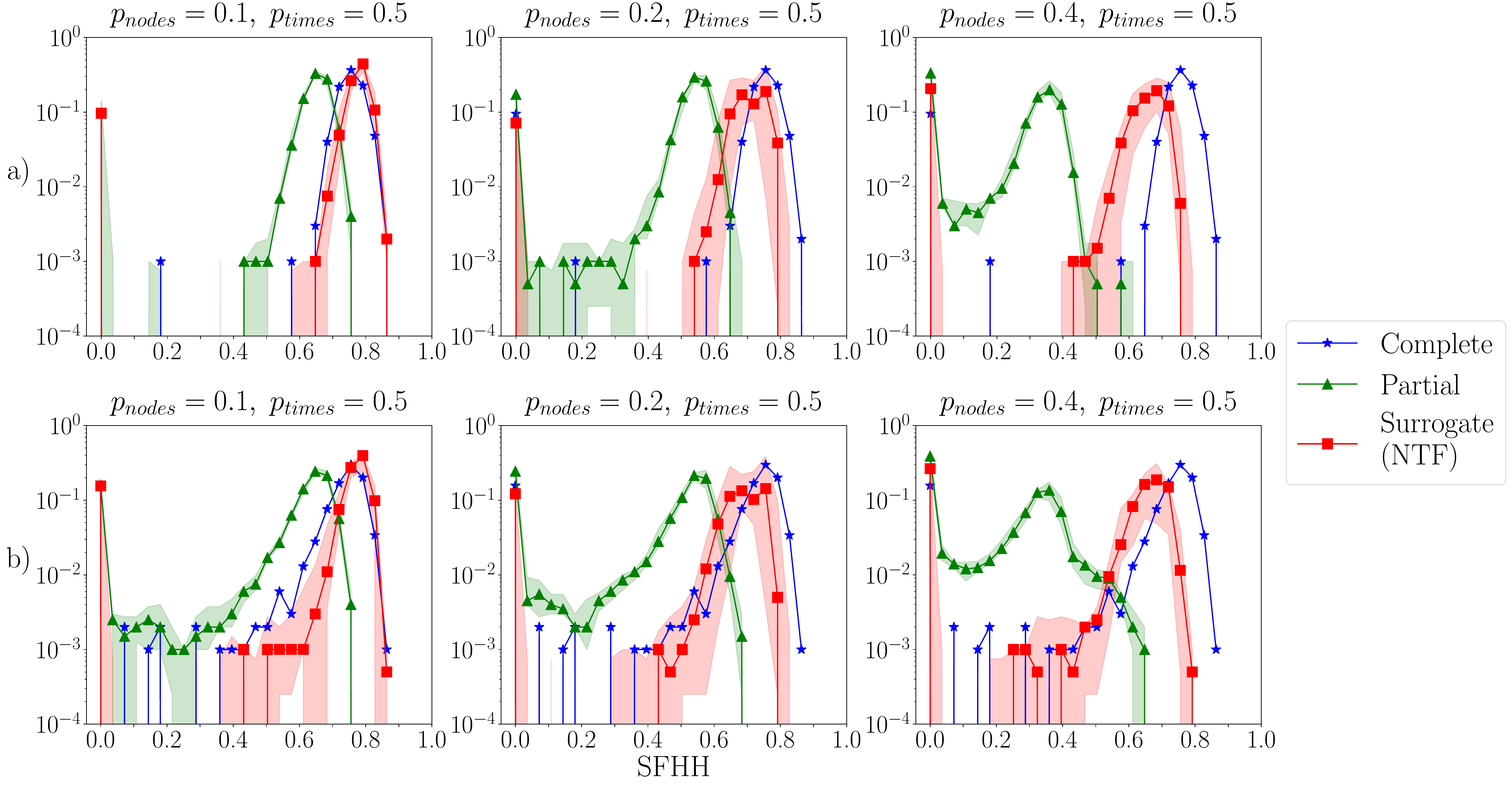}
  \caption{Distributions of epidemic sizes computed in the complete, partial, and surrogate cases for the SFHH dataset. Each panel corresponds to one 
  value of the fraction $p_{nodes}$ of nodes with partial information and one couple of
  spreading parameters: a) $\beta = 0.3$ and $\mu = 0.10$; b) $\beta = 0.25$ and $\mu = 0.08$. 
  For the partial and surrogate cases, the symbols and 
  lines show the median distribution of the epidemic size computed from the results relative to the $10$ different sets of nodes with partial information, while the shaded area is delimited by the $25$-th and $75$-th percentiles.}
\label{sirsfhh}
\end{figure*}

We obtain similar results for the HT09 and SFHH datasets, for which we report the results 
of numerical simulations of the SIR process for various values of the spreading parameters ($\beta$,$\mu$) and of $p_{nodes}$
in Fig.s~\ref{sirht09} and \ref{sirsfhh}. In all cases, using only the 
partial temporal network in the numerical simulations of the SIR process leads to a clear underestimation of the epidemic sizes; this underestimation
becomes worse as $p_{nodes}$ increases. The simulations performed on the surrogate network are systematically much closer to the ones based on the complete
information, leading thus to a strong improvement in the prediction of the epidemic risk.
For the SFHH case however, we observe that the agreement becomes worse 
when $p_{nodes}$ increases. This is 
 due to the fact that during large conferences attendees tend to follow the schedule less rigorously than in small ones
 and move around and engage in contacts more freely and in a more random way, 
 thus leading to less correlated activity patterns. 
 In such a case, auxiliary data could be useful, as in the case we describe in the next section.

\subsubsection{Using both the partial network and an additional proxy}

By construction, the method based on the NTF cannot handle extreme cases such as missing information for all nodes at the same time or 
activity fully missing for some nodes (as considered in \cite{genois2015compensating}). 
To address this limitation, we propose an extension of our method that makes it possible to take advantage of additional information that can be used as a proxy for human proximity. To this aim, we consider the JNTF method described in Sec.~\ref{approxjntf}. We test this extended method on the LSCH dataset, for which we have access to the approximate position of individuals in time. There are indeed $15$ locations in the school: $10$ classes, the cafeteria, the playground, two staircases and a control room. 
The resulting bipartite temporal network relating individuals and locations
is composed by $N = 241$ nodes representing individuals, $15$ nodes representing locations, and $130$ temporal 
snapshots: the tensor representing this additional information has dimensions $I = 241$, $J = 15$ and $K = 130$. Let us notice that due to the temporal resolution selected, a node might appear in several locations in the same snapshot. 

To compare the methods of construction of surrogate data 
based on the JNTF and on the NTF decompositions, we simulated a loss of data on the contact network for a fraction of nodes $p_{nodes} = 0.2$ and for
several fractions of the timespan  $p_{times}=\left[0.6,0.8,1\right]$ selected consecutively on the temporal activity of the nodes. 
Here, the set of nodes with partial activity is the same for all values of $p_{times}$, so that we can compare the outcomes of the SIR process and the impact of 
incrementally removing larger fractions of the temporal activity of the same nodes.
We build surrogate data using the methods based on the NTF and on the JNTF decompositions, performed respectively on the partial contact networks 
and on the joint partial contact and location networks. For the JNTF, we impose a coupling on the first and third dimensions (i.e., we impose 
the equality of the matrices obtained in the decompositions of the two tensors, for $\mathbf{A}$ on the one hand and for $\mathbf{C}$ on the other hand), 
as the two networks have the same nodes related to individuals and the same snapshots in time. 
The choice of this coupling relies on the reasonable assumption that co-located individuals are more likely to be in contact. 

\begin{figure*}
  \includegraphics[width=\textwidth]{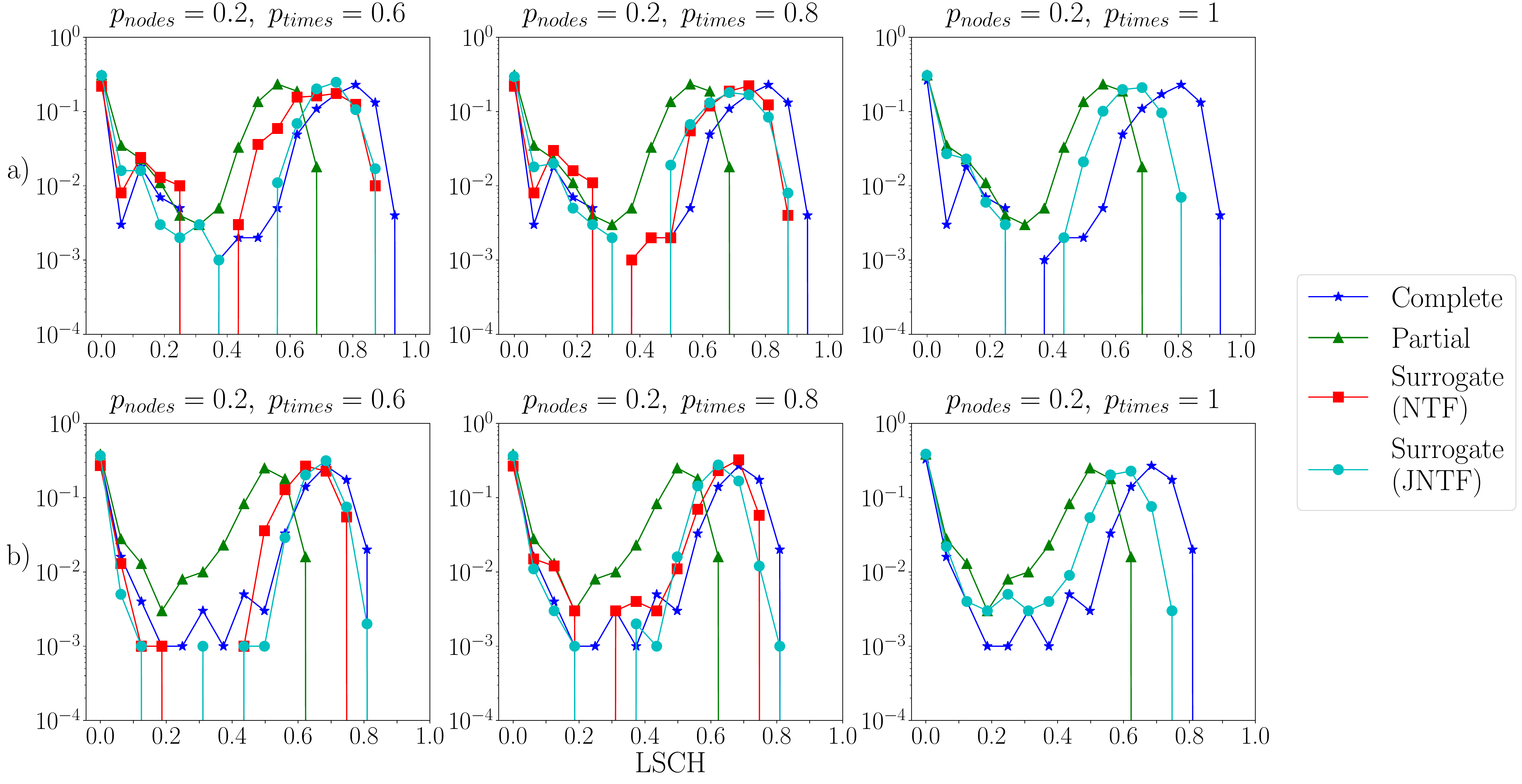}
  \caption{Results achieved by the surrogate construction methods  using either the NTF or the JNTF. 
 Each subfigure corresponds to a fixed fraction of nodes $p_{nodes} = 0.2$ with
 increasing loss of information for these nodes: $p_{times}=0.6,0.8,1$. 
 The JNTF is computed on the two tensors representing the temporal social network of the LSCH dataset and the related temporal location network. 
 Here, we report the epidemic size distributions of the complete, partial and surrogate networks 
  for SIR processes with two different couples of infection/recovery probabilities: a) $\beta = 0.3$ and $\mu = 0.3$, b) $\beta = 0.15$ and $\mu = 0.25$. 
  In the last panels of both a) and b), as $p_{times}=1$ no information can be gained on the nodes
  with missing data using simply NTF, so the results of the NTF-based method coincide with the ones obtained by simulating the SIR over the network with partial information.}
\label{jntftime}
\end{figure*}

 \begin{figure}
  \includegraphics[width=0.5\textwidth]{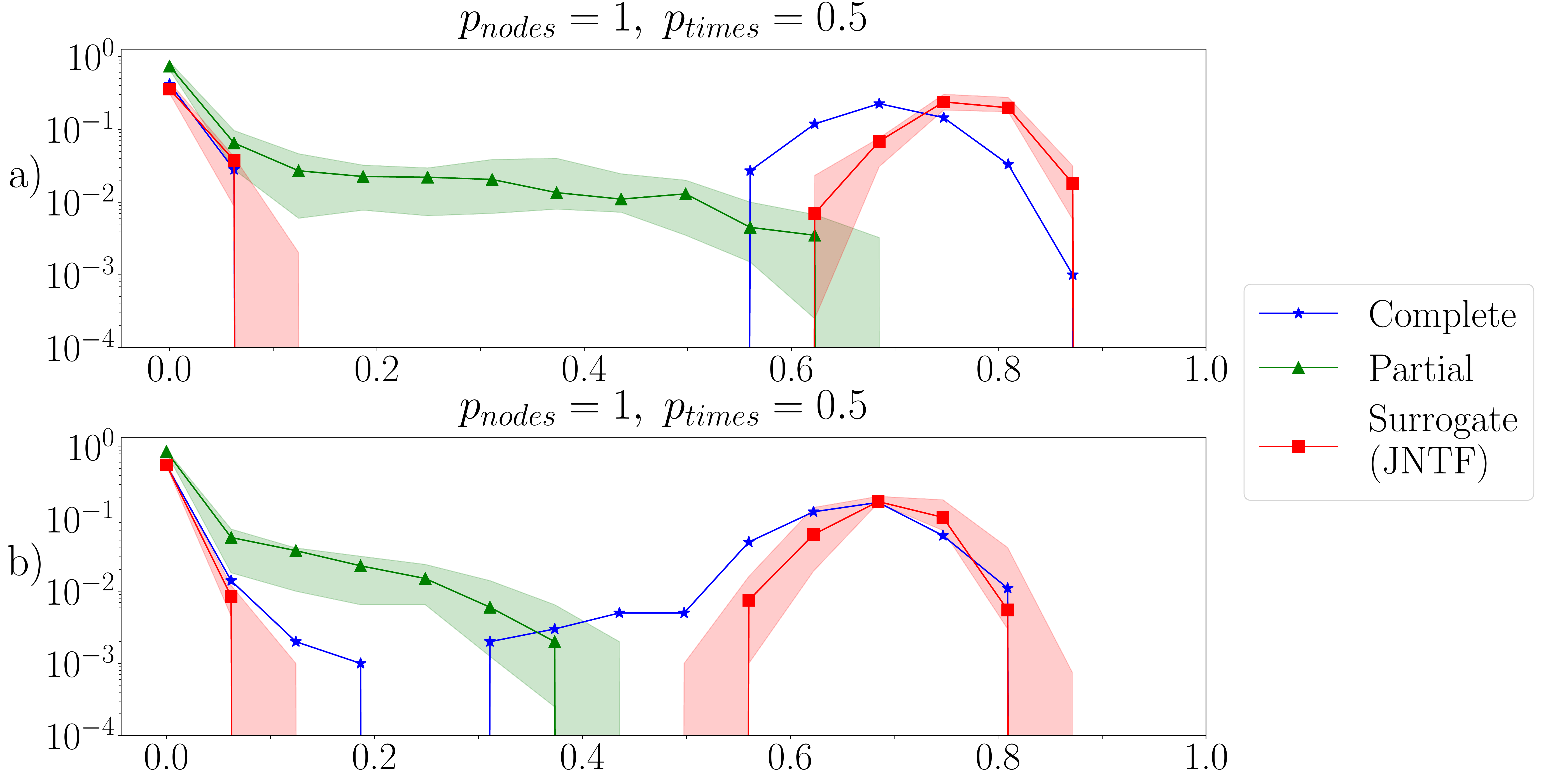}
  \caption{Results achieved by the surrogate construction method using the JNTF. Each subfigure corresponds to the case in which $p_{nodes} = 1$ and $p_{times}=0.5$. The JNTF is computed on the two tensors representing the temporal human proximity network of the LSCH dataset and the related temporal location network. 
  Here, we report the results for two different couples of infection/recovery probabilities: a) $\beta = 0.3$ and $\mu = 0.3$, b) $\beta = 0.15$ and $\mu = 0.25$. 
 The symbols and lines show the median distributions of epidemic sizes computed over the $10$ random generations of the missing times, 
 while the shaded area is delimited by the $25$-th and $75$-th percentiles.}
\label{jntfnode}
\end{figure}

In Fig.~\ref{jntftime}, we display the results obtained for $p_{nodes} = 0.2$ and $p_{times} = 0.6, 0.8,1$. The different subfigures in a) and b) correspond to SIR processes with the following infection and recovery probabilities: $\beta = 0.3$, $\mu = 0.3$, and $\beta = 0.15$ and $\mu = 0.25$. 
As in the previous cases, the distributions of epidemic sizes computed on the network with partial information 
show a clear underestimation of the number of individuals infected. For $p_{times}=0.6$ and $p_{times}=0.8$,
the results achieved with both procedures based on the NTF and the JNTF are in agreement and give a
better estimation of the distributions of epidemic sizes obtained in the original network than using the incomplete data. 
Finally, we report in the rightmost panels of each subfigure the extreme case for which no information about $20\%$ of nodes is 
present in the partial contact network ($p_{times}=1$). 
In this case, as no information is present at all for the selected nodes, no correlated activity pattern concerning them can be inferred by the NTF decomposition,
which thus cannot help recover any information on these nodes. Using JNTF proves then helpful and the surrogate network built using this method, which combines
contact and location information, yields a distribution of epidemic size  closer to the one obtained on the complete network. 
This shows that the external information provided by the approximated location of individuals in the school helps to infer 
possible correlations in the contact activity, even for nodes for which no contact activity was initially known. 
We note however that the epidemics sizes remain underestimated with respect to the complete network. 

Finally, we consider a case in which all nodes had some activity missing. We simulate a scenario in which all the nodes have lost half of their activity:
for each node, we zero out its activity for half of the times taken at random among the times it was active. 
In that way each node has a different set of times during which its activity is erased. We apply the method based on the JNTF in such scenarios with $10$ 
random generations of the missing times for each node. The epidemic size distributions obtained are represented on Fig.~\ref{jntfnode} together with the one measured with the complete network. While the distributions measured on the partial networks strongly differ from those obtained with the complete network, the approach based on the JNTF yields a much better estimation of the real epidemic size distribution.

\section{Discussion}
\label{Con}

In this work, we have proposed a new approach to face the problem of estimating the outcome of spreading processes on temporal human 
proximity networks built from incomplete information. Our method leverages the existent correlations in the observed activity of the nodes 
to recover the contact properties of nodes whose activity is partially missing. To this aim, we rely on tensor decomposition techniques able to 
extract the mesoscale properties of temporal networks. In practice, the methodology we put forward follows two main steps: 
(i) the extraction of structures from the partial network through tensor decomposition 
and (ii) the construction of a surrogate network that is used in numerical simulations to estimate the outcome of spreading processes. 

In the first step we use NTF handling missing values to decompose the partial networks into a sum of components, each describing a mesoscale structure.
In this step, we take into account the fact that the tensor describing the temporal contact network is based on incomplete information, 
and the decomposition is thus performed using only the part of the tensor composed of known contacts or known non-contacts. This leads to
a good approximation of the temporal activity of the nodes with partial information.
In the second step, we determine which links belong to each mesoscale structure, and use the binarized activity timeline of each structure to fill in the unknown 
part of the timeline of each link with missing information.
This is enough to obtain a surrogate network comparable to the empirical one when we use the NTF procedure.
However, the JNTF might  overestimate the activity of the links with missing activity and leads to a distribution of weights
(number of contacts per pair of individuals) that is more homogeneous than in the original case. 
As the heterogeneity properties of contact networks are well known to have a crucial role in determining the outcome of spreading processes,
we adjust the weight distribution of the links for which only partial information was available: this step is made
possible by the robustness under sampling of the contact network weight distribution. We can thus rely on the weight distribution
measured on those links for which no information is missing in the incomplete network to extract at random values and assign them to 
the links with missing information.

We tested our method on three different datasets describing face-to-face contacts between individuals in different contexts (two conferences and a primary school),
represented as temporal human proximity networks, on which we simulated a loss of data, namely a loss of information concerning a fraction
of the individuals, each for a fraction of the total timeline.
We have shown that the method is able to well estimate the outcome of a spreading process, even when half of the activity is missing for
$40\%$ of the nodes, while the epidemic sizes 
are strongly underestimated when we simulate the process on the networks with incomplete information. 

The performance of the method based on the NTF however decreases
when the amount of missing information drastically increases. In particular, it is not applicable if no information at all is available
for some of the nodes. To deal with this issue, we have 
proposed an adaptation of the method based on the joint factorization of multiple tensors (JNTF). 
The JNTF is indeed a natural extension that allows to integrate information encoded in multiple networks. It is particularly adequate
if the information available concerns on the one hand the contacts of individuals and on the other hand their (approximate) location, encoded in a bipartite
temporal network in which a link is drawn between a node representing an individual and a node representing a location 
when the individual is detected in that location. The joint factorization of the tensors representing the partial temporal network of contacts and the temporal network
encoding positions, constrained to extract mesoscale structures with the same nodes and the same activity timelines
in both tensors, can then help to recover the missing information about the contacts.
We have tested this alternative method in a dataset for which both contacts and approximated positions of nodes in time are available.
We have shown that it yields results similar to the NTF when both methods can be applied.
When information about some nodes is completely lost, i.e., when the method based on NTF alone cannot be used, using JNTF allows us 
to recover part of the missing information and yields a distribution of epidemic sizes closer to the original than when using the partial network in the simulations.
It is also the case when as much as $50\%$ of the contact activity of each node has been lost in the data.

The methodology we have presented manages to cope with missing information in various contexts. Importantly, and contrarily to
other methods, it does not rely on the availability of metadata or any knowledge on the structure of the population into groups (such as classes
in a school) as in  \cite{genois2015compensating}: the NTF or JNTF decompositions are indeed able to extract 
the effective structure (both in groups and temporal) of the temporal
network of interactions and to assign each individual to one or multiple mesoscale structures, each with its activity timeline.

Some limitations of our method stem from the tensor decomposition itself. For instance, if 
a whole group of correlated links or nodes is missing in the data, it will not be uncovered by the NTF. In such an unfavorable case,
the JNTF method can compensate the lack of correlated activity in the incomplete contact network by relying on auxiliary data when available. 
Moreover, when the temporal network of contacts lacks structure or 
when the nodes with partial information behave in a random way, i.e., do not exhibit any activity correlation with other nodes,
neither decomposition (NTF nor JNTF) might be able to clearly assign them to any mesoscale structure and thus to determine their activity timeline.
A way to cope with this issue might be to attribute 
some random activity (based for instance on the average behaviour) to those nodes with missing information
that are not found in any mesoscale structure.

Another limitation, which can be easily overcome though, is that for the JNTF we rely on the distribution of weights measured in the partial network.
Indeed, if the remaining information is not representative enough (too much information missing), the heterogeneity properties of the surrogate
network might be affected
as the weights assigned to the links with missing information are taken among those measured. We could however
take advantage of the known robustness of the weight distributions in different contexts \cite{barrat2015face,mastrandrea2016estimate}
to use publicly available weight distributions collected in other contexts.

Finally, a natural direction for future research is the extension of our technique to infer mesoscale properties of human proximity 
networks even when no direct information on contacts is available, but only proxies such as approximate locations of individuals have been collected.

%
%

\section{Methodology}
\label{methods}
Here we describe some technical details regarding the different steps of our procedure.

\subsection{NTF Rank Selection}

The selection of the number of components $R$ for the decomposition is guided by the Core Consistency Diagnostic~\cite{bro2003new}, which estimates to which extent the PARAFAC model 
$\llbracket\boldsymbol\lambda; \mathbf{A}, \mathbf{B}, \mathbf{C}\rrbracket$
with a given rank $r$ (i.e., with a sum of $r$ components) is appropriate to represent the data. The core consistency is a measure that has $100$ as an upper bound and values above $50$ are usually considered acceptable. Here, we computed the core consistency values between the tensor with partial information and its approximation for $r\in\left[2,\dots ,R_{max}\right]$, for $5$ realizations of the optimization procedure starting from different
initial conditions for $\mathbf{A}$, $\mathbf{B}$ and $\mathbf{C}$ for each value of $r$.
We select a rank $R = R_{cc} - 1$, where $R_{cc}$ is the smallest rank for which the core consistency value for each of the 
$5$ realizations is lower than $85$. This threshold is selected to ensure an approximation as faithful as possible to the original tensor. For the JNTF case, we use 
as a hint for the number of components to be selected, the one obtained with the NTF on the temporal contact network with partial information. It is worth noting that,
when the amount of available information in the data decreases, the value of $R$ determined by the core consistency can vary. 
This is due to the fact that by erasing an increasing percentage of node activities the correlated activity patterns are increasingly perturbed and might
be destroyed: a smaller number of components (sub-networks composed by links having a correlated activity in time) is then detected. 

\subsection{JNTF Computation}
\label{jntf}
To integrate data from multiple sources we use the Joint Non-negative Tensor Factorization (JNTF), described in Sec.~\ref{jntf}. To this aim, we adapt 
the  Alternating large-scale Non-negativity-constrained Least Squares (ANLS)
framework and in particular the way to compute the Karush-Kuhn-Tucker (KKT) optimality conditions in a Block Principal Pivoting (BPP) framework~\cite{park2012}. Here, we illustrate the adaptation details to solve Eq.~\eqref{jntfS} in the case of two data sources 
that are coupled in two dimensions (case study in Sec.~\ref{Res}), i.e. $S=2$, 
$\mathbf{A} = \mathbf{A}_1 = \mathbf{A}_2$, and $\mathbf{C} = \mathbf{C}_1 = \mathbf{C}_2$, such that Eq.~\eqref{jntfS} becomes:
\begin{align}
\min &\left( \frac{1}{2}\left\|\mathbfcal{T}_1-\llbracket\boldsymbol\lambda_1;\mathbf{A},\mathbf{B}_1,\mathbf{C}\rrbracket\right\|^2_F + \frac{\alpha_1}{2}\left\|\boldsymbol\lambda_1\right\|^2_2 \right.  \nonumber\\
+&\left. \frac{1}{2}\left\|\mathbfcal{T}_2-\llbracket\boldsymbol\lambda_2;\mathbf{A},\mathbf{B}_2,\mathbf{C}\rrbracket\right\|^2_F + \frac{\alpha_2}{2}\left\|\boldsymbol\lambda_2\right\|^2_2 \right) \label{jntf2s}\\
&\mbox{s.t.} \boldsymbol\lambda_{1,2},\mathbf{A},\mathbf{B}_{1,2},\mathbf{C}\geq 0\;. \nonumber
\end{align} 
Note that we added here the regularization terms $\frac{\alpha_{1,2}}{2}\left\|\boldsymbol\lambda_{1,2}\right\|^2_F$, which are sparsity penalties.

To solve the problem through the BPP algorithm we rewrite the minimization problem as a multiple right hand side problem of the form:
\begin{equation}
\min\left\|\mathbf{V}\mathbf{X}-\mathbf{W}\right\|^2_F\;,
\label{mrhside}
\end{equation}
and solve it for each of the factor matrices (here we include $\boldsymbol\lambda_{1,2}$ in the factor matrices). To this aim, we start by solving the problem for the factor matrices $\mathbf{B}_1$ and $\mathbf{B}_2$ which are respectively present in the first and third term of Eq.~\eqref{jntf2s}. With respect to these factor matrices we can rewrite the equation by using the $2$-mode matricization~\cite{P1} of $\mathbfcal{T}_1$ and $\mathbfcal{T}_2$, which leads to the following approximations:
\begin{align*}
&\mathbf{T}_{1,\left(2\right)}\approx \mathbf{B}_1\boldsymbol\Lambda_1\left(\mathbf{C}\odot\mathbf{A}\right)^T \\
&\mathbf{T}_{2,\left(2\right)}\approx  \mathbf{B}_2\boldsymbol\Lambda_2\left(\mathbf{C}\odot\mathbf{A}\right)^T\;\;,
\end{align*}
where
\begin{align*} &\boldsymbol\Lambda_1=\mbox{diag}\left(\lambda_{1,1},\dots ,\lambda_{1,R}\right) \\ &\boldsymbol\Lambda_2=\mbox{diag}\left(\lambda_{2,1},\dots ,\lambda_{2,R}\right)\;.
\end{align*}
We can rewrite the approximations as 
\begin{align*}
&\mathbf{T}^T_{1,\left(2\right)}\approx\left(\mathbf{C}\odot\mathbf{A}\right)\boldsymbol\Lambda_1\mathbf{B}^T_1 \\
&\mathbf{T}^T_{2,\left(2\right)}\approx\left(\mathbf{C}\odot\mathbf{A}\right)\boldsymbol\Lambda_2\mathbf{B}^T_2\;\;,
\end{align*}
where $\boldsymbol\Lambda^T_{1,2} = \boldsymbol\Lambda_{1,2}$, and thus
\begin{align*}
&\mathbf{B}^T_1\approx\boldsymbol\Lambda^{-1}_1\left(\mathbf{C}\odot\mathbf{A}\right)^{\dagger}\mathbf{T}^T_{1,\left(2\right)} \\
&\mathbf{B}^T_2\approx\boldsymbol\Lambda^{-1}_2\left(\mathbf{C}\odot\mathbf{A}\right)^{\dagger}\mathbf{T}^T_{2,\left(2\right)}\;\;.
\end{align*}
By using the following property of the Khatri-Rao product, for which
\[
\left(\mathbf{C}\odot\mathbf{A}\right)^{\dagger} = \left(\mathbf{C}^T\mathbf{C}\ast\mathbf{A}^T\mathbf{A}\right)^{\dagger}\left(\mathbf{C}\odot\mathbf{A}\right)^T
\]
we can write the approximation as
\begin{align*}
&\boldsymbol\Lambda_1\left(\mathbf{C}^T\mathbf{C}\ast\mathbf{A}^T\mathbf{A}\right)\mathbf{B}^T_1\approx\left(\mathbf{C}\odot\mathbf{A}\right)^T\mathbf{T}^T_{1,\left(2\right)}\;,\\
&\boldsymbol\Lambda_2\left(\mathbf{C}^T\mathbf{C}\ast\mathbf{A}^T\mathbf{A}\right)\mathbf{B}^T_2\approx\left(\mathbf{C}\odot\mathbf{A}\right)^T\mathbf{T}^T_{2,\left(2\right)}\;\;.
\end{align*}
The related subproblems of Eq.~\eqref{jntf2s} for $\mathbf{B}_1$ and $\mathbf{B}_2$ are then reduced to:
\begin{align*}
&\min_{\mathbf{B}_1}\frac{1}{2}\left\|\boldsymbol\Lambda_1\left(\mathbf{C}^T\mathbf{C}\ast\mathbf{A}^T\mathbf{A}\right)\mathbf{B}^T_1-\left(\mathbf{C}\odot\mathbf{A}\right)^T\mathbf{T}^T_{1,\left(2\right)}\right\|^2_F \;,\\
&\min_{\mathbf{B}_2}\frac{1}{2}\left\|\boldsymbol\Lambda_2\left(\mathbf{C}^T\mathbf{C}\ast\mathbf{A}^T\mathbf{A}\right)\mathbf{B}^T_2-\left(\mathbf{C}\odot\mathbf{A}\right)^T\mathbf{T}^T_{2,\left(2\right)}\right\|^2_F\;.
\end{align*}
Finally, the subproblems can be respectively written in the form of Eq.~\eqref{mrhside} by assigning
\begin{align*}
&\mathbf{V} = \boldsymbol\Lambda_1\left(\mathbf{C}^T\mathbf{C}\ast\mathbf{A}^T\mathbf{A}\right)\;,\;\;\mathbf{X} = \mathbf{B}^T_1 \\
&\mbox{and}\;\; \mathbf{W} = \boldsymbol\Lambda_2\left(\mathbf{C}\odot\mathbf{A}\right)^T\mathbf{T}^T_{1,\left(2\right)}\;, \\
&\mathbf{V} = \left(\mathbf{C}^T\mathbf{C}\ast\mathbf{A}^T\mathbf{A}\right)\;,\;\;\mathbf{X} = \mathbf{B}^T_2 \\
&\mbox{and}\;\; \mathbf{W} = \left(\mathbf{C}\odot\mathbf{A}\right)^T\mathbf{X}^T_{2,\left(2\right)}\;.
\end{align*}
The solution to the subproblems is now straightforward as illustrated in~\cite{park2012}. 

We now need to rewrite the subproblems for the factors $\mathbf{A}$ and $\mathbf{C}$ that are present in both first and third terms of Eq.~\eqref{jntf2s}. We show the procedure for $\mathbf{A}$ (which is analogous for $\mathbf{C}$). First, we write the minimization problem by using the $1$-mode matricization of $\mathbfcal{T}_1$ and $\mathbfcal{T}_2$ ($3$-mode for $\mathbf{C}$):
\begin{align*}
\min_{\mathbf{A}}&\left( \frac{1}{2}\left\|\left(\mathbf{C}\odot\mathbf{B}_1\right)\boldsymbol\Lambda_1\mathbf{A}^T-\mathbf{T}^T_{1,\left(1\right)}\right\|^2_F \right. \\
+ &\left. \frac{1}{2}\left\|\left(\mathbf{C}\odot\mathbf{B}_2\right)\boldsymbol\Lambda_2\mathbf{A}^T-\mathbf{T}^T_{2,\left(1\right)}\right\|^2_F\; \right).
\end{align*}
By following the procedure shown above, we can write the problem in the following way:
\begin{align*}
\min_{\mathbf{A}}&\left( \frac{1}{2}\left\|\boldsymbol\Lambda_1\left(\mathbf{C}^T\mathbf{C}\ast\mathbf{B}^T_1\mathbf{B}_1\right)\mathbf{A}^T-\left(\mathbf{C}\odot\mathbf{B}_1\right)^T\mathbf{T}^T_{1,\left(1\right)}\right\|^2_F + \right. \\
+ &\left. \frac{1}{2}\left\|\boldsymbol\Lambda_2\left(\mathbf{C}^T\mathbf{C}\ast\mathbf{B}^T_2\mathbf{B}_2\right)\mathbf{A}^T-\left(\mathbf{C}\odot\mathbf{B}_2\right)^T\mathbf{T}^T_{2,\left(1\right)}\right\|^2_F\; \right),
\end{align*}
in which we can assign
\begin{align*}
&\mathbf{V}_1 = \boldsymbol\Lambda_1\mathbf{C}^T\mathbf{C}+\mathbf{B}^T_1\mathbf{B}_1\;, \mathbf{W}_1 = \left(\mathbf{C}\odot\mathbf{B}_1\right)^T\mathbf{T}^T_{1,\left(1\right)}\;,\\
&\mathbf{V}_2 = \boldsymbol\Lambda_2\mathbf{C}^T\mathbf{C}+\mathbf{B}^T_2\mathbf{B}_2\;, \mathbf{W}_2 = \left(\mathbf{C}\odot\mathbf{B}_2\right)^T\mathbf{T}^T_{2,\left(1\right)}\;, \\
&\mbox{and}\;\;\mathbf{X} = \mathbf{A}^T\;,
\end{align*}
leading to
\begin{equation}
f\left(\mathbf{X}\right) = \min_{\mathbf{X}}\left(
\frac{1}{2}\left\|\mathbf{V}_1\mathbf{X}-\mathbf{W}_1\right\|^2_F + \frac{1}{2}\left\|\mathbf{V}_2\mathbf{X}-\mathbf{W}_2\right\|^2_F \right)\;.
\label{kktnew}
\end{equation}
To solve the problem in Eq.~\eqref{kktnew} we need to adapt the KKT conditions, which result to be
\begin{align*}
&\nabla f\left(\mathbf{X}\right) = \underbrace{\left(\mathbf{V}^T_1\mathbf{V}_1+\mathbf{V}^T_2\mathbf{V}_2\right)}_{\mathbf{V}_{1,2}}\mathbf{X}-\underbrace{\left(\mathbf{V}^T_1\mathbf{W}_1 + \mathbf{V}^T_2\mathbf{W}_2\right)}_{\mathbf{W}_{1,2}} \\
&\nabla f\left(\mathbf{X}\right)\geq 0\;,\;\nabla f\left(\mathbf{X}\right)^T\mathbf{X}=0\;,\;\mathbf{X}\geq 0\;,
\end{align*}
whose solution is given by solving 
\[
\mathbf{X}^T\mathbf{V}^T_{1,2}-\mathbf{W}^T_{1,2} = 0\;.
\]
Since Eq.~\eqref{jntf2s} includes regularization terms, we have to adapt the KKT conditions for the minimization problem with respect to $\boldsymbol\lambda_1$ and $\boldsymbol\lambda_2$. We show the procedure for $\boldsymbol\lambda_1$, which is analogous for $\boldsymbol\lambda_2$. We consider the cost function built from the terms in which $\boldsymbol\lambda_1$ is involved (the first and second in Eq.\eqref{jntf2s}):
\[
f_{\boldsymbol\lambda_1} = \frac{1}{2}\left\|\mathbfcal{X}_1-\llbracket \boldsymbol\lambda_1; \mathbf{A},\mathbf{B}_1,\mathbf{C}\rrbracket\right\|^2_F + \frac{\alpha_1}{2}\left\|\boldsymbol\lambda_1\right\|^2_2\;,
\]
and we rewrite it through the vectorization:
\[
f_{\boldsymbol\lambda_1} = \frac{1}{2}\left\|\mbox{vec}\left(\mathbf{C}\odot\mathbf{B}\odot\mathbf{A}\right)\boldsymbol\lambda_1-\mbox{vec}\left(\mathbfcal{X}\right)\right\|^2_F + \frac{\alpha_1}{2}\left\|\boldsymbol\lambda_1\right\|^2_2\;.
\]
Minimizing the cost function $f_{\boldsymbol\lambda_1}$ is equivalent to minimize $f'_{\boldsymbol\lambda_1}$, obtained by incorporating the regularization term as follows:
\[
f'_{\boldsymbol\lambda_1} = \frac{1}{2}\left\|\underbrace{\binom{\mbox{vec}\left(\mathbf{C}\odot\mathbf{B}_1\odot\mathbf{A}\right)}{\sqrt{\alpha_1}}}_{\mathbf{V}}\underbrace{\boldsymbol\lambda_1}_{\mathbf{x}}-\underbrace{\binom{\mbox{vec}\left(\mathbfcal{X}\right)}{0}}_{\mathbf{w}}\right\|^2_F\;;
\]
The solution to the minimization of $f'_{\boldsymbol\lambda_1}$ follows from~\cite{park2012}. 

\subsection{Otsu Method}

The Otsu method~\cite{fang2009study} is commonly used in image processing to recover the different levels of grey in pixels. The same idea can be used on the temporal activities of the components by thinking of them as images composed by $K$ pixels of different values (level of activation). In particular, the method for $2$-dimensional functions assumes that the function given as an input contains values that follow a bimodal distribution. The method computes then the optimal threshold, defined by minimizing the intra-class variance and by maximizing the inter-class variance. The resulting optimal threshold can be used to divide the values of the function into two groups. 

Here, we use the threshold given by the Otsu method to define whenever a component is active or not. This is done by applying the Otsu method on the temporal activity of each component. The values above the threshold correspond to the temporal activation of the component, while the values below the threshold correspond to the times in which the component is inactive. 

\subsection{SIR processes}
\label{sir}
To simulate how an infectious disease propagates in a population, and thus how the related dynamical process spreads over the network, we run an SIR process over the network. The SIR model assumes that each individual in the population can be in one of three states: Susceptible, Infectious, Recovered. The propagation of the disease starts from a single individual, who is in the Infectious state (the seed of the process) while all others are initially susceptible. At each timestep, 
each susceptible individual in contact with an infectious one has a probability $\beta$ to become infectious. Each
infectious individual recovers with probability $\mu$ per timestep.
For simulation purposes, the first node to be infected is chosen among the nodes that are not in the set of those with partial information. 
For each network and each set of values of the parameters $\beta$ and $\mu$,
we run $1000$ simulations of the spreading process. In particular, 
the parameter space we consider a priori is given by all the couples of probabilities $\left(\beta,\mu\right)$, with $\beta ,\mu$ in the range $\left[0.001,1\right]$. 
We select the suitable couples of parameters according to the conditions described in Sec.~\ref{outSIR} and we show in the main text the results corresponding to
two representative examples of parameter couples for each dataset. Results related to other couples of parameters are similar to the ones
shown in the figures.

\section*{Supplementary information}

\subsection*{Approximated network}
\label{Sappnet}
\begin{figure*}
\includegraphics[width=\textwidth,height=17cm]{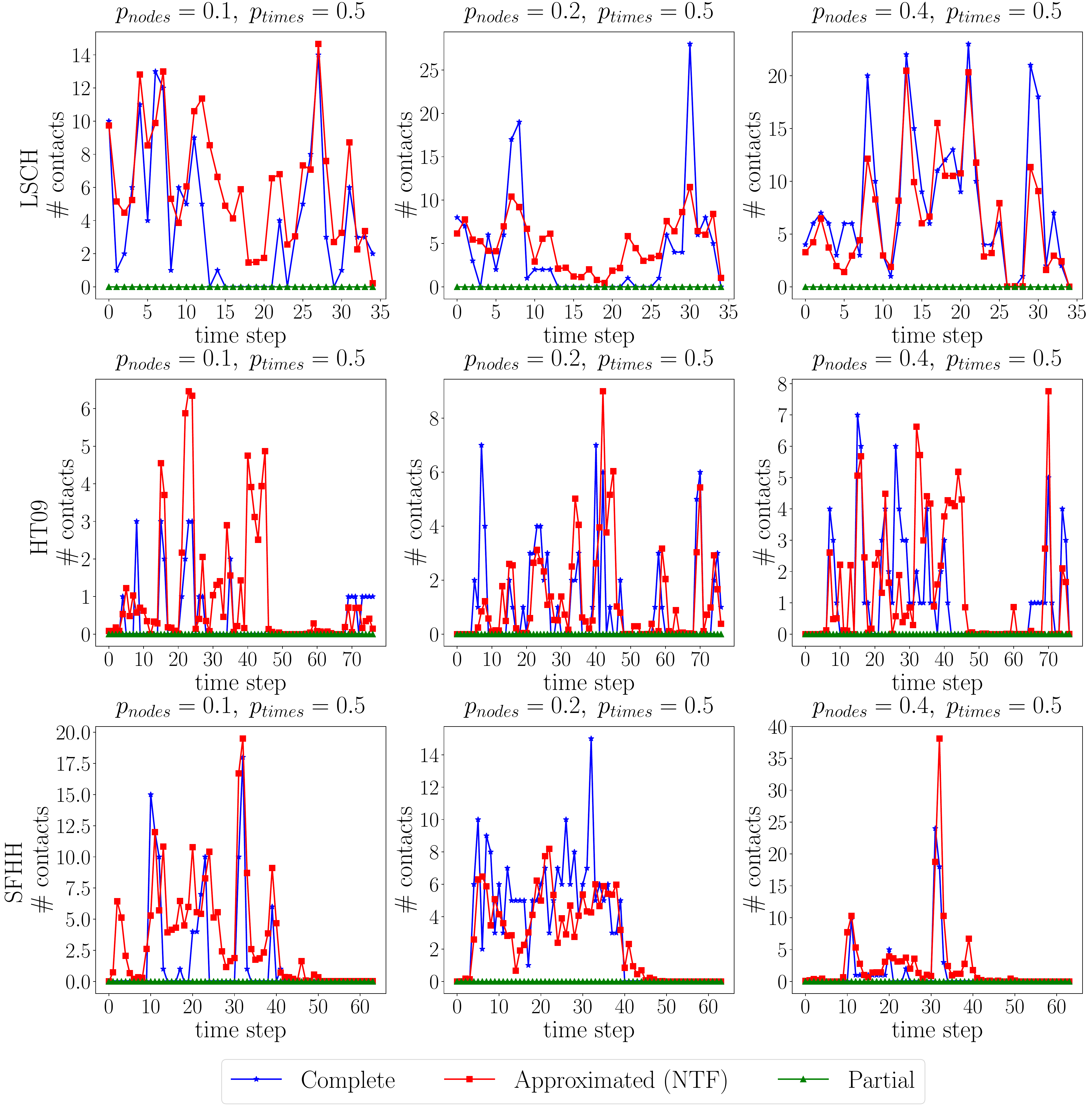}
\caption{\textbf{Temporal activity}: example of the total number of contacts in time of a node for each dataset and case study. 
We report the temporal activity in the complete network, in the one with partial information, and in the one obtained approximated by the NTF method. 
We show here the temporal activity only for the time steps in which we lost the information about the contacts of the node, corresponding
in our scenario to the first half of the timeline.}
 \label{fig:corr}
\end{figure*}

By decomposing a network affected by missing information through the Non-negative Tensor Factorization (NTF) we are able to reconstruct some of the structural and temporal properties observed in the complete network. One of these characteristics is the overall contact activity of the nodes. In particular, given a node whose information is partially missing, we are able to reconstruct its overall activity in time, i.e., the number of contacts related to that node at each time. We have indeed shown
that the activities in the complete network and in the approximated one are significantly correlated and have high Pearson correlation coefficients. 
In Fig.~\ref{fig:corr}, we report a representative example for the overall temporal activity of a single node in the complete, partial, and approximated network for each dataset: LSCH, HT09, and SFHH. As we consider a scenario in which the missing activity concerns, in each case, the first half of the timeline, 
the activity of the nodes with missing information is $0$ in the partial data, while it shows non-trivial patterns in the complete network, which are
partially recovered in the approximated network.

\subsection*{SIR on the approximated network}
\label{SSIRappnet}
\begin{figure*}
\includegraphics[width=\textwidth]{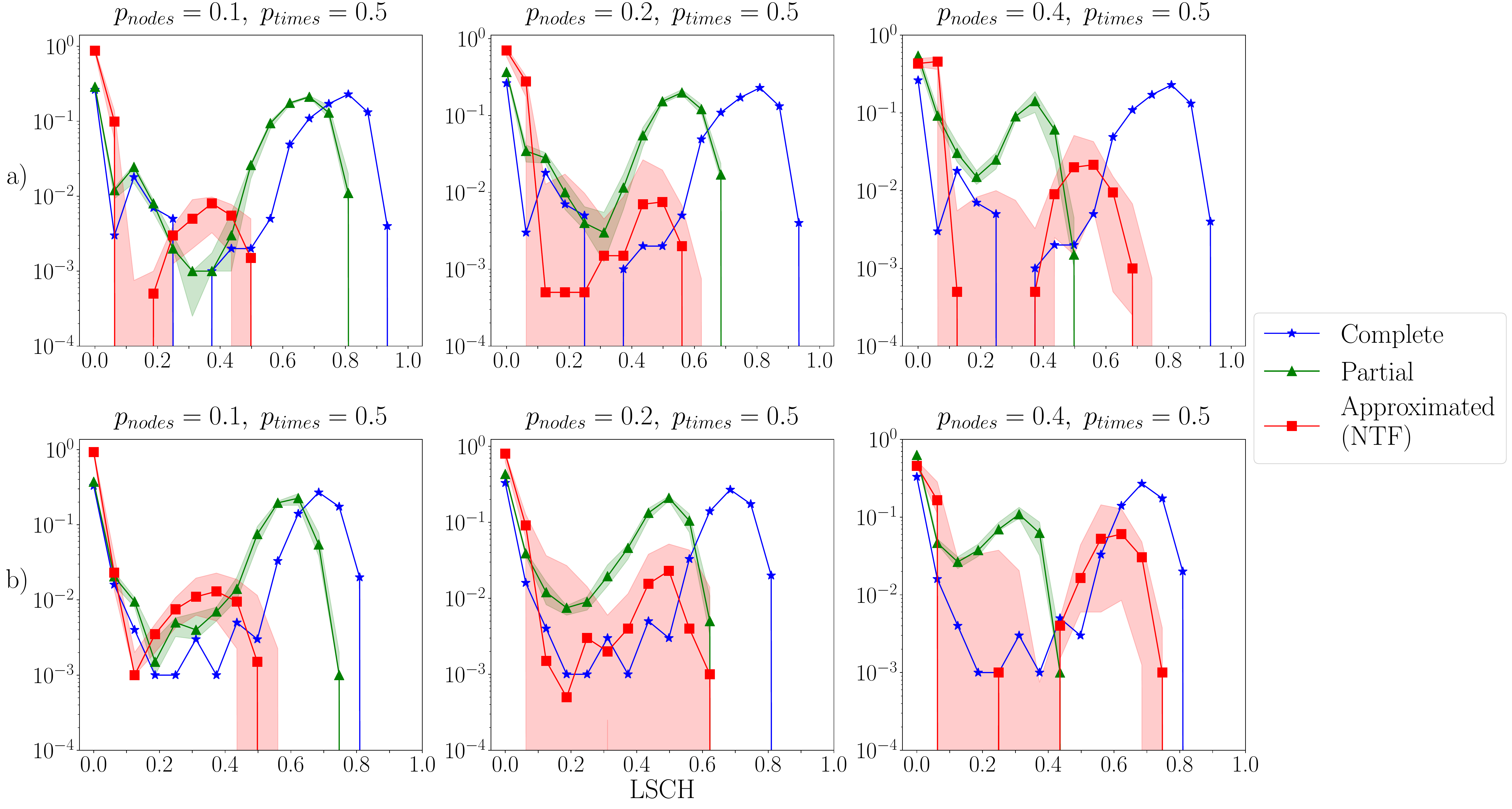}
\caption{\textbf{Outcome of an SIR process}: over the complete network, the partial network and the network approximated through the NTF for the LSCH dataset. Each panel corresponds to one fraction $p_{nodes}$ of nodes with partial information and one couple of spreading parameters: a) $\beta = 0.30$ and $\mu = 0.30$; b) $\beta = 0.15$ and $\mu = 0.25$. The lines show the median distribution of the epidemic size computed from the results relative to  $10$ different sets of nodes
with missing information, while the shaded area is delimited by the $25$-th and $75$-th percentiles.}
 \label{sir_inter}
\end{figure*}

We have shown that by approximating a network with partial information via NTF we are able to recover some of its properties, such as the overall temporal activity of the nodes. However, as discussed in the main text, this information is not enough to obtain outcomes of a spreading process 
close to the ones obtained on the complete network, because the approximation tends to make the distributions of weights more homogeneous than the empirical one. 
We illustrate this fact by showing in Fig.~\ref{sir_inter} the distributions of epidemic sizes of an SIR process simulated on the complete, partial
and approximated networks for the LSCH dataset, for different spreading parameters ($\beta = 0.30$ and $\mu = 0.30$ and $\beta = 0.15$ and $\mu = 0.25$), 
different fractions of nodes missing ($0.1,0.2,0.4$) and fraction of times with missing data $p_{times} = 0.5$.

 Here, we reported the results only for the LSCH dataset, as it is the one characterized by highly correlated activity patterns and thus the one for which the approximations obtained through the NTF are closer to the complete case. In the primary school indeed, students' activity is determined by the school schedule and they are divided in classes. This makes their activity highly correlated as their contacts are more homogeneous during the daily class activities. However, as we can see from Fig.~\ref{sir_inter}, even if the approximated network is close to the complete one in terms of the overall activity patterns displayed, this is not the case for the outcome of the spreading process. The epidemic sizes in the case of the partial and approximated networks are far from the complete network case. This strong underestimation is due by the fact that when we are approximating the network via NTF, the method tends to approximate the activity patterns of the nodes in the same component as fully correlated, thus connecting the nodes in the same component and rendering the network more homogeneous.

\subsection*{Weight Distribution}
\label{Swd}

\begin{figure*}
\includegraphics[width=\textwidth]{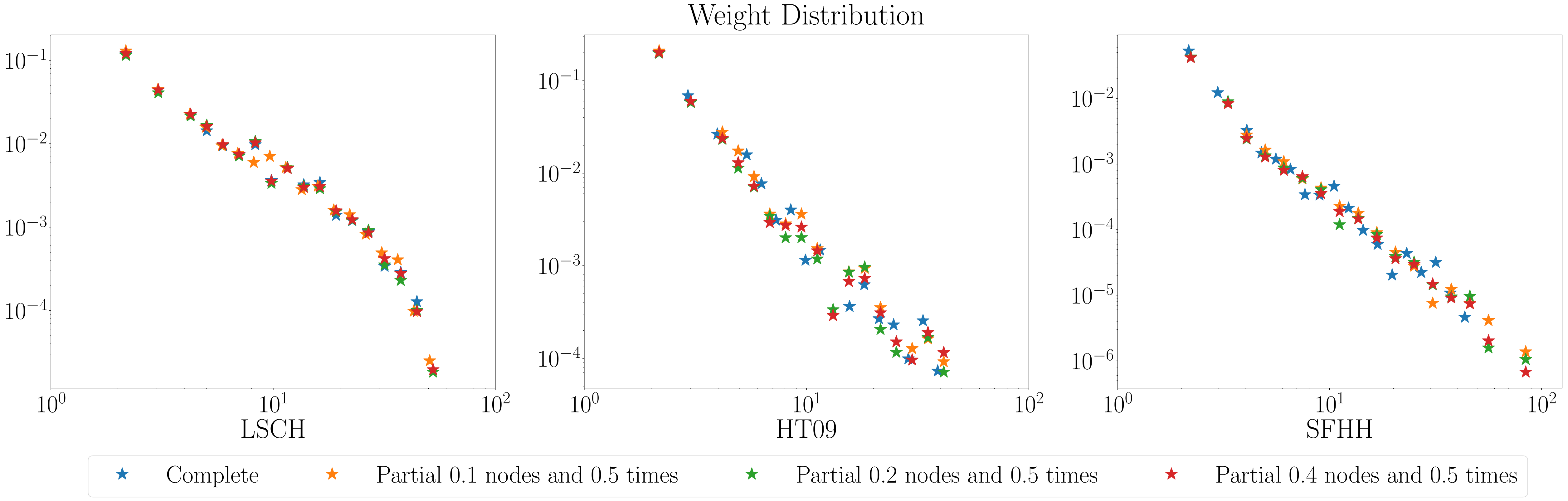}
\caption{\textbf{Weight distributions} in the three datasets for the complete network and the partial network with increasing percentage of nodes with missing
information: $p_{nodes}\in\left[0.1,0.2,0.4\right]$ and $p_{times}=0.5$.}
 \label{wd}
\end{figure*}

For the JNTF case, 
in order to obtain outcomes of SIR processes close to the case of the complete network, we have to reintroduce the heterogeneity properties 
into the approximated network, and use the resulting surrogate network
to simulate the SIR process. We reintroduce the heterogeneity properties by reassigning the weights, i.e., the total number of contacts, to the links 
involving nodes with partial information. In particular, we measure the weight distribution on the available part of the partial network. 
Then, we use this distribution to pick at random a weight that will be reassigned to the link 
whose activity was approximated via the JNTF. We rely on such a process as the weight distribution of a network is robust to various sampling procedures,
as shown in Fig.~\ref{wd} for the sampling considered in this paper: 
for each dataset, we compare  the weight distribution of the complete network with the corresponding weight distributions computed on 
the links of partial networks not involving any nodes with missing information, with $p_{nodes}\in\left[0.1,0.2,0.4\right]$. By construction, these distributions do not depend on $p_{times}$. The results clearly show that
the weight distributions measured in the partial and complete networks are consistent, meaning that,
even in the case in which we miss almost half of the links in the partial network, we are able to compute a weight distribution similar to the
one of the complete network.

\end{document}